\documentclass[aps,preprint]{revtex4-1} 
\usepackage{graphicx}
\usepackage{amssymb}
\usepackage{epstopdf}
\usepackage{amsmath}
\usepackage{bm}

\newcommand{\dd}{\partial}
\newcommand{\eps}{\epsilon}

\newcommand\refeq[1]{(\ref{#1})}
\newcommand\reffig[1]{Figure~\ref{#1}}
\newcommand\of[1]{\left( #1 \right)}

\newcommand\mat[1]{{\mathbb #1}}

\begin{document}

\title{The Dynamics of the Schr\"odinger-Newton System with Self-Field Coupling}

\author{J. Franklin}
\email{jfrankli@reed.edu}
\author{Y. Guo}
\author{K. Cole Newton}

\affiliation{Department of Physics, Reed College, Portland, Oregon 97202,  USA}

\author{M. Schlosshauer}

\affiliation{Department of Physics, University of Portland, Portland, Oregon 97203,
USA}

\begin{abstract}
We probe the dynamics of a modified form of the Schr\"odinger-Newton system of gravity coupled to single particle quantum mechanics.  At the masses of interest here, the ones associated with the onset of ``collapse" (where the gravitational attraction is competitive with the quantum mechanical dissipation), we show that the Schr\"odinger ground state energies match the Dirac ones with an error of $\sim 10\%$.  At the Planck mass scale, we predict the critical mass at which a potential collapse could occur for the self-coupled gravitational case, $m \approx 3.3$ Planck mass, and show that gravitational attraction opposes Gaussian spreading at around this value, which is a factor of two higher than the one predicted (and verified) for the Schr\"odinger-Newton system.  Unlike the Schr\"odinger-Newton dynamics, we do not find that the self-coupled case tends to decay towards its ground state; there is no collapse in this case.
\end{abstract}
\maketitle

\section{Introduction}

In a recent paper, we studied the spectrum of a modified form of the usual Schr\"odinger-Newton system (SN) of gravity coupled to quantum mechanics (SN was originally developed in~\cite{Bon}).  
Now we turn to the spherical dynamics of the self-coupled gravity introduced, in this quantum mechanical setting, in~\cite{us}.

For the SN system, we have Newtonian gravity determining the potential $\Phi$ using the wave function itself to describe the mass density, so the coupled system is
\begin{equation}\label{SN}
\begin{aligned}
i \, \hbar \, \frac{\dd \Psi}{\dd t} &= -\frac{\hbar^2}{2 \, m} \, \nabla^2 \Psi + m \, \Phi \, \Psi \\
\nabla^2 \Phi &= 4 \, \pi \, G \, m \, \Psi^* \, \Psi.
\end{aligned}
\end{equation}
The spectrum and dynamics of this system of equations has been studied extensively, and its relevance to single-particle collapse similarly explored -- see~\cite{Carlip, Giulini2} and references therein for a review of that discussion.

Motivated by the special relativistic notion that energy and mass are equivalent, we modified the gravitational piece to include the self-gravity of $\Phi$ itself -- the resulting static theory of gravity was originally introduced by Einstein in~\cite{Einstein}, and has been re-developed periodically (see~\cite{FandN, DandH, GiuliniSC, FranklinAJP}, for example).  When we combine this new gravity model with Schr\"odinger's equation, we get 
\begin{equation}\label{SCSG}
\begin{aligned}
i \, \hbar \, \frac{\dd \Psi}{\dd t} &= -\frac{\hbar^2}{2 \, m} \, \nabla^2 \Psi + m \, \Phi \, \Psi \\
\nabla^2 \sqrt{\Phi} &= \frac{2 \, \pi \, G}{c^2} \, m \, \Psi^* \, \Psi \, \sqrt{\Phi}.
\end{aligned}
\end{equation}
Here, we have modified the field equation for gravity to reflect the same sort of self-consistent self-coupling that is found in full general relativity (albeit in a scalar setting).  The form comes from considering the combined gravity/quantum mechanical equation, from~\cite{Moller, Rosenfeld},
\begin{equation}\label{MRG}
G_{\mu\nu} = 8 \, \pi \, \langle \hat T_{\mu\nu} \rangle,
\end{equation}
and making a gravitational field equation in~\refeq{SCSG} that is more like the nonlinear (Einstein tensor) left-hand side of~\refeq{MRG} than the linear Poisson equation for gravity found in~\refeq{SN}.  Both SN and our modification take the source to be $m \, \Psi^*\, \Psi$, and the approach can be viewed either as part of a multi-body Hartree approximation, or fundamental (the many-body view would not change the gravitational field equation here -- we would still have to incorporate the energy self-coupling).  In this work, we will take a single-particle wave-function which cannot be viewed, by itself, as a Hartree approximation (due to the lack of self-interaction in the Hartree approach~\cite{Adler}).  There are other ways of extending the gravitational field equation to capture additional relativistic effects, like introducing the gravito-magnetic contribution as in~\cite{Manfredi}.  That allows the ``magnetic" component of weak-field gravity to play a role in the SN setting.  But that extension retains the linearity of the gravitational field equations themselves.  We are working in a complementary direction, in which we extend to include the self-energy coupling that leads to the nonlinearity of general relativity.

The dynamics of the SN system, in particular, the details of spherical collapse, have been studied, and our goal is to compare the SN collapse with the (potential) spherical collapse of an initial Gaussian evolved using~\refeq{SCSG}.  
En route to that comparison, we will first consider the role of the relativistic Dirac equation with the modified gravity.  Then we will estimate the critical mass at which the gravitational interaction balances the spreading of a free Gaussian, for both SN and the modified gravitational form.  In the SN case, this critical mass defines the boundary between collapse (to a ground state) and dissipation.  For the self-coupled case, there is no collapse to the ground state, although at the critical mass, there is a balance between gravity and quantum mechanical dissipation.

\section{Dirac Equation}

Given that we are using the relativistic notion of energy and mass equivalence to motivate the use of the modified form of gravity appearing in~\refeq{SCSG}, it is reasonable to introduce the competing relativistic effects on the quantum mechanical side.  If we start with the Dirac Lagrangian, coupled to the Lagrangian appropriate to the modified form of gravity (that gravitational Lagrangian can be found in~\cite{us,FranklinAJP}), 
\begin{equation}
\mathcal L = i \, \hbar \, \bar \Psi \, \gamma^\nu \, \dd_\mu \, \Psi - m \, c^2 \, \bar \Psi \, \Psi - m \, \Phi \, \bar \Psi \, \gamma^0 \, \Psi - \frac{c^2}{8 \, \pi \, G \, \Phi} \nabla \Phi \cdot \nabla \Phi,
\end{equation}
then the resulting Dirac equation and modified gravity coupling gives an eigenvalue problem for the ground state that looks like (already in spherical coordinates):
\begin{equation}
\begin{aligned}
\left[ \begin{array}{cc} m \, c^2 + m \, \Phi & \hbar \, c \, \of{-\frac{d}{d r} + \frac{\kappa}{r}} \\ \hbar \, c \, \of{\frac{d}{dr} + \frac{\kappa}{r}}  & - m\, c^2 + m \, \Phi \end{array} \right] \,
\left[\begin{array}{c} u \\ v \end{array} \right] &= E \, \left[\begin{array}{c} u \\ v \end{array} \right] \\
\frac{d^2}{d r^2} \, \of{r \, \sqrt{\Phi}} &= \frac{2 \, G \, m}{c^2 \, r} \, \of{u^* u + v^* v} \, \sqrt{\Phi},
\end{aligned}
\end{equation}
where we take $\kappa = 1/2$ (no orbital angular momentum).

\begin{figure}[htbp] 
   \centering
   \includegraphics[width=4in]{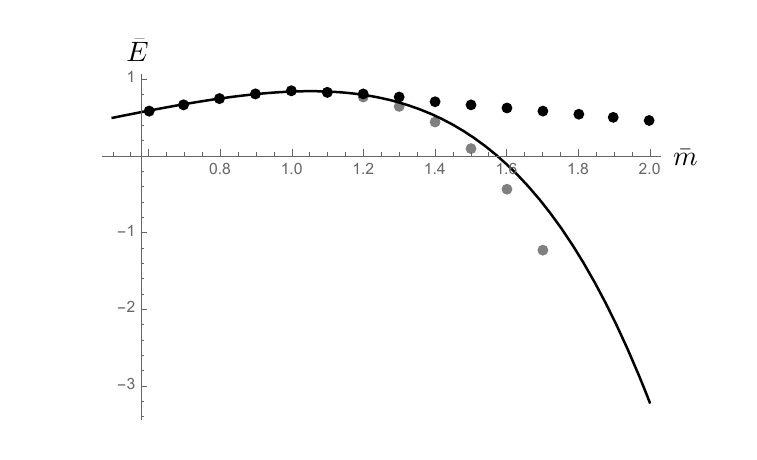} 
   \caption{The (dimensionless) energy, as a function of mass (in units of Planck mass), for the ground state of the modified-gravity-Dirac system is shown with black dots.  The same calculation using a Newtonian gravitational field and the Dirac equation is shown in gray dots, and the solid line is the SN ground state energy, for comparison.}
   \label{fig:west}
\end{figure}

We can solve this coupled system just as we did in~\cite{us} -- the numerical method doesn't change significantly, although there are relativistic details that need to be addressed (the presence of negative energy states, for example, means we need to be careful how we identify the ground state).  We modified our method to accommodate the additional complexity, and proceeded to find the ground state energies for the new system (see~\cite{DanGuo}).  The Dirac ground state energy, as a function of mass, is shown in~\reffig{fig:west}.  In that figure, we also show the effect of using the Dirac equation together with Newtonian gravity, and the ground state energy of SN itself, all for comparison.

\begin{figure}[htbp] 
   \centering
   \includegraphics[width=4in]{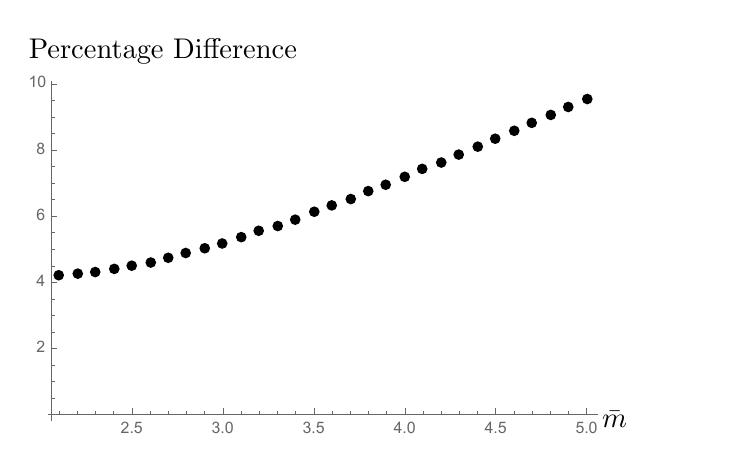} 
   \caption{The percentage difference between the ground state energies as computed using the Dirac equation and the Schr\"odinger equation.  Mass is in units of Planck mass.}
   \label{fig:comparE}
\end{figure}

By how much does the ground state energy change when we use the Dirac equation instead of Schr\"odinger?  We can compare the energy estimates directly, as shown in~\reffig{fig:comparE}.  There, the percentage difference between the energies computed using the Schr\"odinger equation vs. the Dirac equation are shown (both cases use the modified form of gravity, of course).
The divergence of the two energies at the masses shown is relatively mild, with a difference of $10\%$ at five Planck masses.

For the temporal evolution of an initial Gaussian, we'll use the Schr\"odinger equation, where the numerical method is easy to generate and verify.  We will work with large masses, between $1$ and $5$ Planck mass, where the ground state energies differ by $\sim 5-10 \%$ between Schr\"odinger and Dirac.  While we are well within the relativistic regime at these masses, the difference in energy is small, and we expect the basic qualitative dynamics to hold using the Schr\"odinger equation instead of the Dirac equation.  

\section{Dimensionless Form, Units}

Starting from~\refeq{SN} and~\refeq{SCSG}, let $P \equiv r\, \Psi$, and then set $r = r_0 \, R$, $t = t_0 \, T$, and let $\Phi = c^2\, \bar \Phi$, $P = P_0 \, \bar P$, and $m = m_0 \, \bar m$ with $m_0 \equiv \sqrt{\frac{\hbar \, c}{G}}$ the Planck mass.  The Schr\"odinger equation becomes
\begin{equation}
-\frac{\dd^2 \bar P}{\dd R^2} + \bar m^2 \, \bar \phi \, \bar P = i \, \bar m \, \frac{\dd \bar P}{\dd T}
\end{equation}
and then we use either Newtonian gravity (top) or the self-coupled form (bottom):
\begin{equation}
\begin{aligned}
\frac{\dd^2}{\dd R^2} \, \of{R \, \bar \phi} &= \frac{\bar m}{R} \, \bar P^* \, \bar P \\
\frac{\dd^2}{\dd R^2} \, \of{R \, \sqrt{\bar \phi}} &= \frac{1}{2} \, \bar m \, \frac{\sqrt{\bar \phi}}{R} \,  \bar P^* \, \bar P 
\end{aligned}
\end{equation}
where we have set
\begin{equation}\label{scales}
r_0 = \frac{\hbar}{\sqrt{2} \, m_0 \, c} \, \, \, \, \, \, \, \, \, \, \, t_0 = \frac{\hbar}{m_0 \, c^2} \, \, \, \, \, \, \, \, \, \, \, P_0 = \frac{c}{\sqrt{4 \, \pi \, m_0 \, G}},
\end{equation}
and $r_0$ is (up to the factor of $1/\sqrt{2}$) the Planck length.

While the SN set has well-known scaling properties (see~\cite{Moroz,Giulinibound}) that allow a single numerical result to be relevant for a wide variety of mass and length scales, the nonlinearity introduced in the self-coupled form of gravity spoils the scaling, so that the numerical results refer only to the mass/length scales used.  We know that the self-coupled scalar gravity reduces to Newtonian gravity for small masses, so the results of previous work will hold at those relevant mass scales (around $10^{10}$ u, for example).  Our goal is to probe the higher mass regime, in which the relativistic correction provided by the self-coupling of the gravitational field is significant, and these scales are basically forced upon us numerically -- the choices in~\refeq{scales} uniquely render the gravitational field equation with unit coefficients.

We'll start with a spherically symmetric Gaussian wave function:
\begin{equation}\label{Psinitial}
\Psi(r,0) = \of{ \pi \, a^2 }^{-3/4} \, e^{-r^2/(2 \, a^2)}
\end{equation}
where $a^2$ is the variance (up to constants) of the initial distribution.  Then our initial, dimensionless $\bar P$ is
\begin{equation}\label{barPinitial}
\bar P(R,0) = r_0 \, R \, \Psi(r_0 \, R,0)/ P_0 = 2 \, \of{\frac{2}{\pi}}^{1/4} \, A^{-3/2} \, R \, e^{-R^2/(2\, A^2)}
\end{equation}
with $a = r_0\, A$.  The normalization of the wave function, in the dimensionless setting, is
\begin{equation}\label{normit}
\int_0^\infty \bar P^* \, \bar P \, d R = \frac{1}{4 \, \pi \, P_0^2 \, r_0} = \sqrt{2}.
\end{equation}

For our initial Gaussians, we will take $a = r_0$, so that $A = 1$.  While we can make $A$ larger to spread out the initial distribution of mass as a source for gravity, there is no natural multiple of $r_0$ to use -- one might try to extend the distribution beyond, for example, its Schwarzschild radius (at $2 \, \sqrt{2} \, \bar m$ in these dimensionless units) -- but then the mass required to achieve collapse also increases, and the initial distribution ends up inside the Schwarzschild radius again~\cite{NOTE1}.  In order to compare with potential experiments, the relevant scale is $a = .5 \times 10^{-6}$ m (as in~\cite{Salzman}), but in our units, this leads to $A \sim 4 \times 10^{28}$, inappropriately large for numerical work.  At the low densities implied by taking $a = .5\, \mu$ m, we know that the predictions of the self-coupled form of gravity match the Newtonian case.  Choosing $A = 1$ allows us to probe the regime in which Newtonian gravity must be augmented by the self-gravity of the field (and additional, as yet unknown, physics). 


\section{Numerical Method}

The collapse dynamics of SN have been studied in~\cite{Giulinibound, HarrisonN, Meter, Salzman}, and all use similar methods to time-evolve initial Gaussians: some variant of Crank-Nicolson and a solver for the gravitational Poisson problem in iterative combination.  Our method is similar, when applied to SN, although we use Verlet to find the gravitational field (as opposed to quadrature or a pseudo-spectral method).  Verlet is easy to apply to the nonlinearity present in the self-coupled gravitational field equation, with its more complicated boundary conditions.  The pieces (Crank-Nicolson and Verlet) can be described separately, but then an iterative step must be involved to achieve a self-consistent solution.  We start by discretizing in space and time via $R_j = j \, \Delta R$ and $T_n = n \, \Delta T$ for constant spacings $\Delta R$, $\Delta T$.  We'll call the value of $\bar P$ (at location $R_j$ and time $T_n$) $\bar P(R_j, T_n) \equiv \bar P^n_j$, and similarly $\bar \phi(R_j, T_n) \equiv \bar \phi^n_j$.

The forward-Euler discretization in time, for the Schr\"odinger piece, reads
\begin{equation}\label{fed}
\bar P^{n+1}_j =\bar P^n_j  -\frac{i}{\bar m} \, \Delta T \, \left[ -\frac{\bar P^{n}_{j+1} - 2 \, \bar P^n_j + \bar P^n_{j-1}}{\Delta R^2} + \bar m^2 \, \bar \phi^n_j \, \bar P^n_j \right].
\end{equation}
This equation holds for all grid points, and we understand that at $j = 0$, we have $\bar P^n_0 = 0$ for all $n$, that's the boundary condition at the origin (for $\Psi$ finite at the origin, as it should be, $P = r\, \Psi$ will be zero at the origin).  The spatial grid will extend to $R_\infty = N \, \Delta R$ for integer $N$,  our choice of numerical infinity, and out there we'll again set $\bar P^n_{N+1} = 0$; the wave function should vanish.

Let the vector $\bar{\bf P}^n$ contain the (unknown) spatial values at time level $n$:
\begin{equation}
\bar{\bf P}^n \dot = \left( \begin{array}{c} P^n_1 \\ P^n_2 \\ \vdots \\ P^n_{N} \end{array} \right).
\end{equation}
and similarly for the vector $\bar{\bm \phi}^n$.  Then we can write the forward Euler discretization (together with the boundary conditions) in terms of a matrix-vector multiplication:
\begin{equation}
\bar{\bm P}^{n+1} = \of{\mat I - i\, \Delta T \, \mat H(\bar{\bm \phi}^n)} \, \bar{\bm P}^n
\end{equation}
where $\mat I$ is the identity matrix, $\mat H(\bar{\bm \phi}^n)$ is defined by~\refeq{fed}, and we highlight its dependence on the gravitational potential.

The backwards Euler version of the problem is 
\begin{equation}
\of{\mat I + i\, \Delta T \, \mat H(\bar{\bm \phi}^{n+1}) } \, \bar{\bf P}^{n+1} = \bar{\bf P}^n,
\end{equation}
and then the Crank-Nicolson method is defined by
\begin{equation}\label{CN}
\of{\mat I + i\, \frac{\Delta T}{2} \, \mat H(\bar{\bm \phi}^{n+1}) } \, \bar{\bf P}^{n+1} = \of{\mat I - i\, \frac{\Delta T}{2} \, \mat H(\bar{\bm \phi}^n)} \, \bar{\bf P}^n.
\end{equation}

For the gravitational field portion, we'll use Verlet, although the details will change slightly between the two forms of gravity for reasons that will become clear as we go.  For Newtonian gravity, we start at ``spatial infinity" (out at $R_N$) with the Newtonian limiting form: $\bar \phi^n_{N} = 1 - \sqrt{2}/R_N$ and $\bar \phi^n_{N+1} = 1 - \sqrt{2}/R_{N+1}$ -- the constant term provides a constant offset ($c^2$ when units are introduced) that doesn't effect the probability density here, but we introduce it for comparison with the modified gravity.  Starting at $N$, we move inwards according to the Verlet update:
\begin{equation}
\bar{\phi}^{n+1}_{j-1} =\frac{1}{R_{j-1}} \, \of{ 2 \, \bar \phi^{n+1}_j  \, R_j - \bar \phi^{n+1}_{j+1} \, R_{j+1}+ \Delta R^2 \, \of{ \frac{\bar m}{R_j} \, \left\vert P^{n+1}_j \right\vert^2 }}.
\end{equation}

The procedure for modified gravity is a little different -- at spatial infinity, we know that Newtonian gravity, for a spherically symmetric source of mass $m$, must limit to $-\frac{G \, m}{r}$ (or $c^2 - \frac{G \, m}{r}$ if a constant offset is desired).  But for the modified gravitational field, we have $c^2 - \frac{G \, \tilde m}{r}$ as the leading contribution at spatial infinity -- the $c^2$ is required so that the modified solutions become Newtonian in the non-relativistic limit (see~\cite{FranklinAJP}), and the ``mass" $\tilde m$ depends on the details of the central distribution (for example, a point mass $m$ at the origin and a sphere of homogeneous mass density and total mass $m$, lead to different values for $\tilde m$).  Since the central distribution of mass will change here, the value for $\tilde m$ is a function of time, a complication we'd like to avoid.

Instead, we'll focus on the value of the field as $r \rightarrow 0$.  For a sphere with homogenous mass density, the internal field $\Phi(r)$ looks like (see~\cite{GiuliniSC,FranklinAJP})
\begin{equation}
\Phi = \left[ \frac{c}{\cosh(R/r_0)} \, \frac{\sinh(r/r_0)}{r/r_0}\right]^2,
\end{equation}
where $R$ is the radius of the sphere and $r_0$ is a constant related to the mass.  As $r \rightarrow 0$, $\phi$ goes to a constant bounded by $c^2$, and the derivative of $\phi$ goes to zero.  
Since we expect there to be some non-zero density near the origin, these are reasonable boundary conditions for our numerical solution, i.e.\ $\bar \phi^n_0 = C$ a constant $\in [0,1]$ and $\bar \phi^n_1 = C$, so that the numerical derivative is approximately zero.  We will pick $C$ so that $\bar\phi^n_N = 1$, its limiting value, at spatial infinity (the best we can do here) by shooting -- i.e.\ we will run forward Verlet:
 \begin{equation}
\sqrt{\bar{\phi}^{n+1}_{j+1}} =\frac{1}{R_{j+1}} \, \left[ 2 \, \sqrt{\bar \phi^{n+1}_j }  \, R_j - \sqrt{\bar \phi^{n+1}_{j-1}} \, R_{j-1}+ \Delta R^2 \, \of{ \frac{\bar m}{2 \, R_j} \, \left\vert P^{n+1}_j \right\vert^2 \, \sqrt{\bar \phi^{n+1}_j} } \right].
\end{equation}
for different values of $C= \sqrt{\phi^{n+1}_0} = \sqrt{\phi^{n+1}_1}$ until $\bar \phi^n_N \approx 1$, using bisection to determine $C$ accurately.

In both of these cases, Newtonian  and modified, we must iterate at each time level to achieve a self-consistent solution -- notice that the left-hand side of~\refeq{CN} depends on $\bar {\bm \phi}^{n+1}$, which we can only get once $\bar{\bf P}^{n+1}$ is known -- but we can't {\it find} $\bar{\bf P}^{n+1}$ without $\bar{\bm \phi}^{n+1}$.  To break out of the recursion, we will define an iterative index $k$ -- let 
$\ ^k\bar{\bf P}^{n+1}$ and $\ ^k\bar{\bm \phi}^{n+1}$ be the $k$ iteration at time-level $n+1$.  For $k = 0$, we define $\ ^0\bar {\bf P}^{n+1} = \bar{\bf P}^n$ and $\ ^0\bar{\bm \phi}^{n+1} = \bar{\bm \phi}^n$.  Now, at level $k$, we update (using the Newtonian update for simplicity) according to:
\begin{equation}
\begin{aligned}
\of{\mat I + i\, \frac{\Delta T}{2} \, \mat H(\ ^{k}\bar{\bm \phi}^{n+1}) } \, \ ^{k+1} \bar{\bf P}^{n+1} &= \of{\mat I - i\, \frac{\Delta T}{2} \, \mat H(\bar{\bm \phi}^n)} \, \bar{\bf P}^n \\
\ ^{k+1}\bar{\phi}^{n+1}_{j-1} &=\frac{1}{R_{j-1}} \, \of{ 2 \, \ ^{k+1} \bar \phi^{n+1}_j  \, R_j - \ ^{k+1} \bar \phi^{n+1}_{j+1} \, R_{j+1}+ \Delta R^2 \, \of{ \frac{\bar m}{R_j} \, \left\vert \ ^{k+1}P^{n+1}_j \right\vert^2 }}
\end{aligned}
\end{equation}
where the top line defines the new value for the wave function, and the second line updates the gravitational field.  We proceed with this iteration until 
\begin{equation}
\| \ ^{k+1}\bar{\bm P}^{n+1} - \ ^k \bar{\bm P}^{n+1} \| < \eps
\end{equation}
where $\eps$ is given -- i.e.\ we continue to iterate until the wave function has stopped changing significantly.  Once we have achieved (numerical) convergence, we set $\bar{\bm P}^{n+1} = \ ^{k+1} \bar{\bm P}^{n+1}$ and $\bar{\bm \phi}^{n+1} = \ ^{k+1} \bar{\bm \phi}^{n+1}$, and we're ready to move on to the next time step.

\section{Critical Mass Estimate}

The goal of this section is to establish mass values for which the behavior of the initial Gaussian shifts from ``mainly quantum", with the initial Gaussian spreading out over time, to ``mainly gravitational", with the initial Gaussian becoming more localized.  One simple way to estimate this mass, from~\cite{Giulinibound}, is to take the free particle solution for the initial Gaussian, which is:
\begin{equation}
\Psi(r,t) = \of{\pi \, a^2}^{-3/4} \, \of{1 + \frac{i \, \hbar \, t}{m\, a^2}}^{-3/2} \, e^{-\frac{r^2}{2 \, a^2 \, \of{1 + \frac{i \, \hbar \, t}{m \, a^2}}}}
\end{equation}
and note that the peak of $r^2 \, \Psi^*(r,t) \, \Psi(r,t)$ is located at
\begin{equation}
r_p(t) = \sqrt{a^2 + \of{\frac{\hbar \, t}{a \, m}}^2}.
\end{equation}
With no gravitational component, $\ddot r_p(0) = \frac{\hbar^2}{a^3 \, m^2}$, the initial acceleration of the most-likely position depends only on $m$ (and the initial variance).  With a gravitational force in place, we have:
\begin{equation}\label{masstimate}
\ddot r_p(0) + \of{-\frac{d}{dr} \, \Phi(r_p(0))} = a_{\hbox{\tiny{net}}}(0),
\end{equation}  
where $ a_{\hbox{\tiny{net}}}(0)$ is the net acceleration (treating the most-likely position as the particle position), and we could arrange to have $ a_{\hbox{\tiny{net}}}(0) = 0$ by taking:
\begin{equation}\label{cusp}
\ddot r_p(0) = \frac{d}{dr} \, \Phi(r_p(0)).
\end{equation}

The $\Phi(r)$ that we use depends on both our choice to consider Newtonian or self-coupled gravity, and the $\rho$ that we decide to use to approximate the initial distribution of ``mass" (in~\cite{Giulinibound}, for example, a point particle at the origin is used to perform this estimate~\cite{NOTE2}).  Since we have a Gaussian profile, we can take $\rho = m \, \Psi^* \, \Psi$ for the initial $\Psi$ given in~\refeq{Psinitial} and use that to solve for $\Phi(r)$.  For Newtonian gravity, the field associated with this source is
\begin{equation}
\Phi(r) = -\frac{G \, m}{r} \, \hbox{erf}\of{\frac{r}{a}},
\end{equation}
and using this in~\refeq{cusp} with $r = a$ (the initial value) gives
\begin{equation}
\frac{h^2}{a^3 \, m^2} + \frac{2 \, G \, m}{a^2 \, e \, \pi} = \frac{G \, m}{a^2} \, \hbox{erfc}(1).
\end{equation}
Since we've taken $a = r_0$, we have $a = \frac{\hbar^2}{\sqrt{2} \, G \, m_0^3}$ (in terms of the Planck mass $m_0$), and we can get rid of $\hbar$ using the Planck mass definition, $\hbar = \frac{G\, m_0^2}{c}$; then the solution to this equation is
\begin{equation}
m = \frac{2^{1/6}}{1 - \frac{2}{e \, \pi} - \hbox{erfc}(1)} \, m_0 \approx 1.5 \, m_0.
\end{equation}

\begin{figure}[htbp] 
   \centering
   \includegraphics[width=3in]{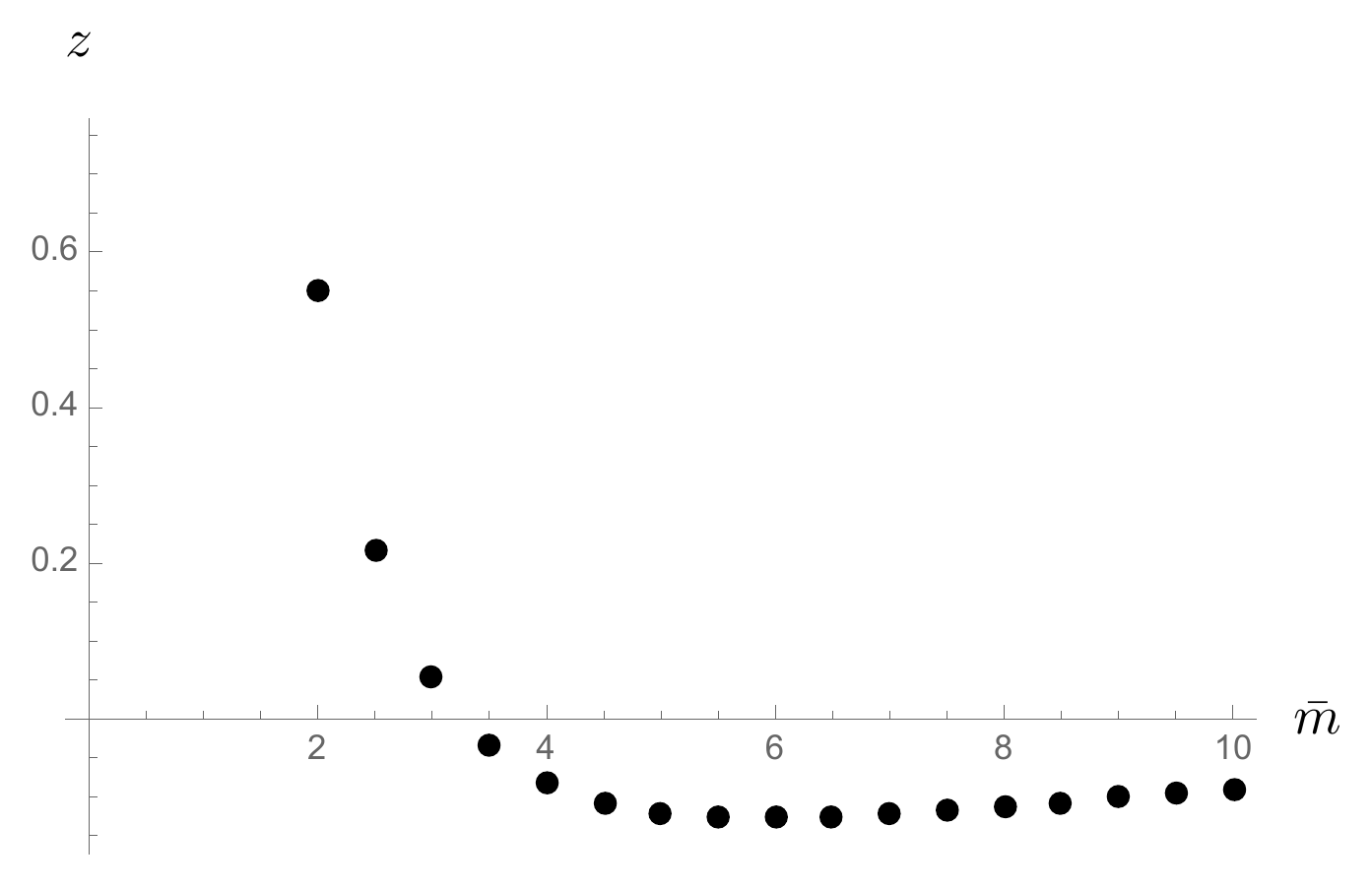} 
   \caption{The dimensionless numerical acceleration, $z$, from~\refeq{zdef}, as function of $\bar m$.}
   \label{fig:zplot}
\end{figure}

For the modified form of gravity, we cannot find $\Phi(r)$ explicitly, so we turn to a numerical approach.  Given the numerical parameters we will use below, we compute the $\bar{\bm \Phi}$ from the initial source (the dimensionless $\bar m \, \bar P^* \, \bar P$ with $\bar P$ and $A = 1$ from~\refeq{barPinitial}, projected onto our numerical grid) using the Verlet method described in Section IV, then approximate the derivative using finite difference (suitably dimensionless, which throws in a factor of $2$) and evaluate that at $\bar r = 1$ ($a$ in our dimensionless units), we subtract $\frac{4}{\bar m^2}$ (the dimensionless form of $\ddot r_p(0)$ here) and then find $\bar m$ such that the difference is close to zero (to within $\eps = 10^{-5}$).  A plot of the difference:
\begin{equation}\label{zdef}
z \equiv \frac{4}{\bar m^2} -2 \,  \frac{\bar \phi_{p+1} - \bar \phi_{p-1}}{2 \, \Delta R}
\end{equation}
with $p \, \Delta R \approx 1$ is shown in~\reffig{fig:zplot}, where we can see that the root lies in between $\bar m =3$ and $4$.  A bisection of $z$ gives $\bar m \approx 3.3$ as the mass associated with the onset of contracting behavior.

\section{Numerical Dynamics}

The numerical results agree well with the predictions from above.   In all cases, we take $N = 1000$ spatial steps, with $R_\infty = 50$, and set $\Delta T = 0.1$.  We can plot the probability densities as functions of time, for the $n=1$, $50$ and $100$ steps to see what sort of evolution is happening.  Following~\cite{Giulinibound}, we also plot the radius in which $90\%$ of the probability lies, this ``$R_{90}(T)"$ value allows us to track the general evolution in time.  We will plot that together with the value associated with a free Gaussian, so we can see what effect gravity (in its various forms) has.  We can further characterize the dynamics by calculating the overlap of the wave function with the ground state (calculated using the methods of~\cite{us}) as a function of time.

The Crank-Nicolson method we use here is not obviously norm-preserving, unlike the original one.  That lack of manifest norm preservation comes from the time-dependence of the matrix operator $\mat H$ appearing on the left and right sides of~\refeq{CN}.  Yet in practice, the norm is preserved well in all the runs, with the maximum difference between the numerical norm and $\sqrt{2}$ (the appropriate normalization from~\refeq{normit}) on the order of $10^{-13}$.

\begin{figure}[htbp] 
   \centering
   \includegraphics[width=4in]{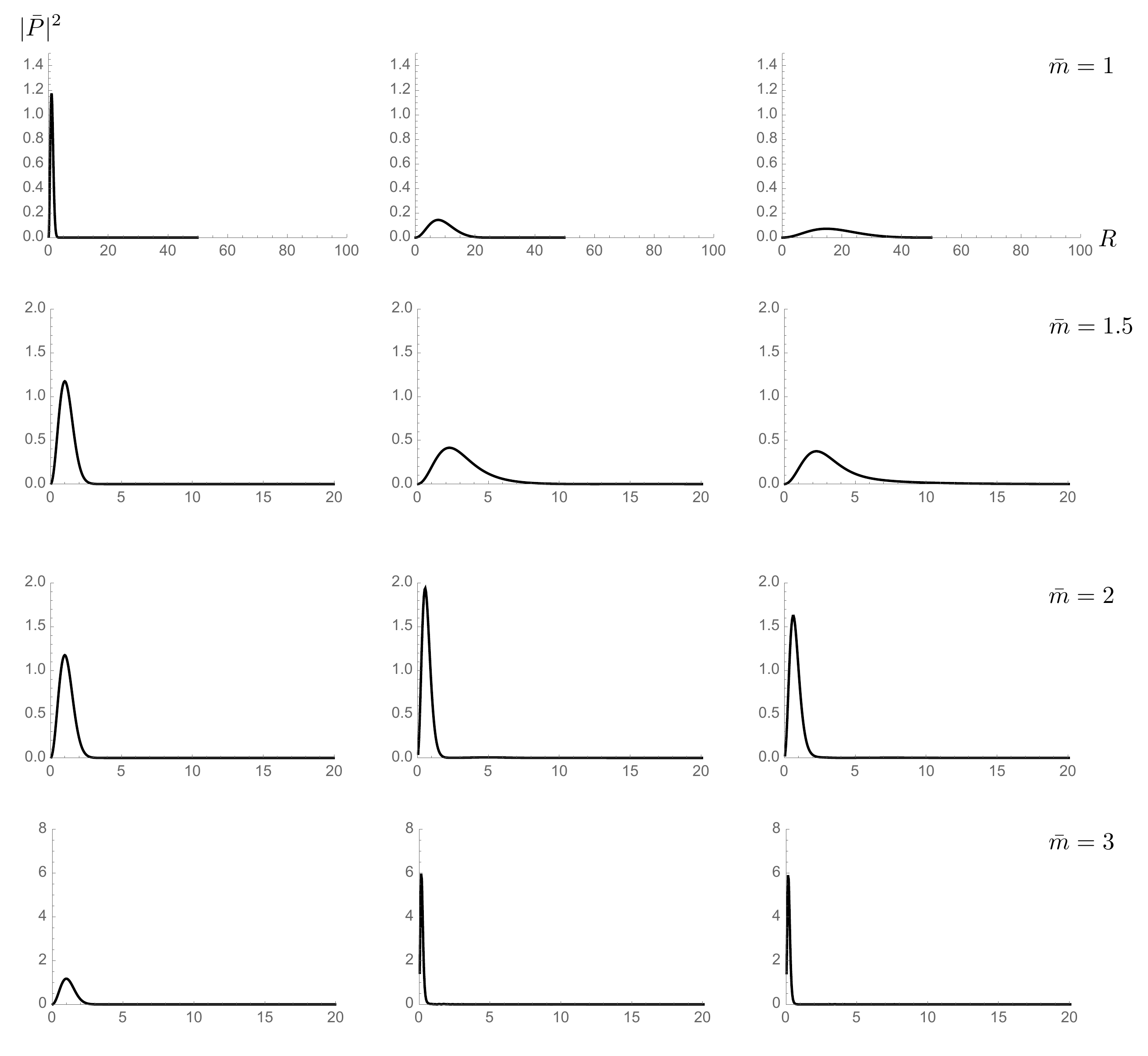} 
   \caption{Probability density as a function of position for SN masses $\bar m = 1$, $1.5$, $2$, and $3$.  Snapshots are shown at $T = 1\, \Delta T$, $50 \, \Delta T$ and $100\, \Delta T$  (left to right) in each case.}
   \label{fig:SNmasses}
\end{figure}

\begin{figure}[htbp] 
   \centering
   \includegraphics[width=4in]{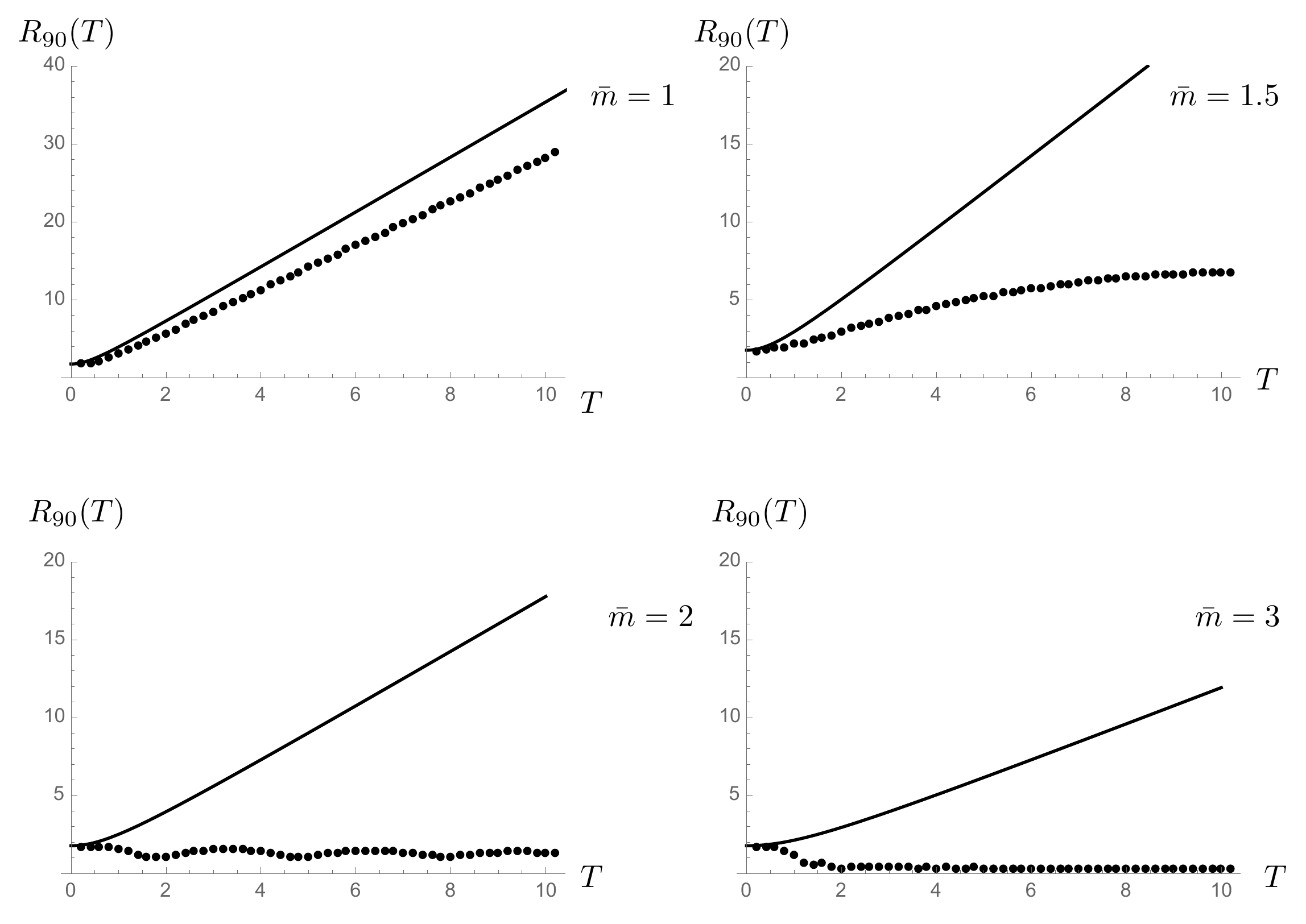} 
   \caption{The values of $R_{90}(T)$ for SN at $\bar m =  1$, $1.5$, $2$ and $3$ are shown as points.  The line is the $R_{90}(T)$ for a free Gaussian.}
   \label{fig:SN90}
\end{figure}

For SN, the probabilities are shown in~\reffig{fig:SNmasses} for $\bar m = 1$, $1.5$, $2$ and $3$, and a plot of $R_{90}(T)$ for each case is shown in~\reffig{fig:SN90}.  There are four different behaviors shown in the plots of $R_{90}(T)$: 1.\ for $\bar m = 1$, the Gaussian spreads out, 2.\ for $\bar m = 1.5$, the Gaussian is oscillating, but with peak position that is further from the origin than at time $T = 0$, 3.\ $\bar m = 2$ has an oscillating solution, where the peak gets closer to the origin and then comes back out and 4.\ a collapse (with minimal oscillation) for $\bar m = 3$ (and greater).  From these plots, the critical mass is somewhere between $1.5$ and $2$, since at $1.5$ we have oscillation above the initial value of $R_{90}(0)$, and at $2$ the oscillation occurs with values less than the initial $R_{90}(0)$.  This estimate of the critical mass basically agrees with our prediction from the previous section, where we found the critical mass to be $\sim 1.5$.
 In~\reffig{fig:SN90}, the solid line shows the value of $R_{90}(T)$ for a free Gaussian (of appropriate mass) for comparison.  As expected, the gravitational coupling makes the spreading behavior slow down compared to the free particle case. 
 
\begin{figure}[htbp] 
   \centering
   \includegraphics[width=4in]{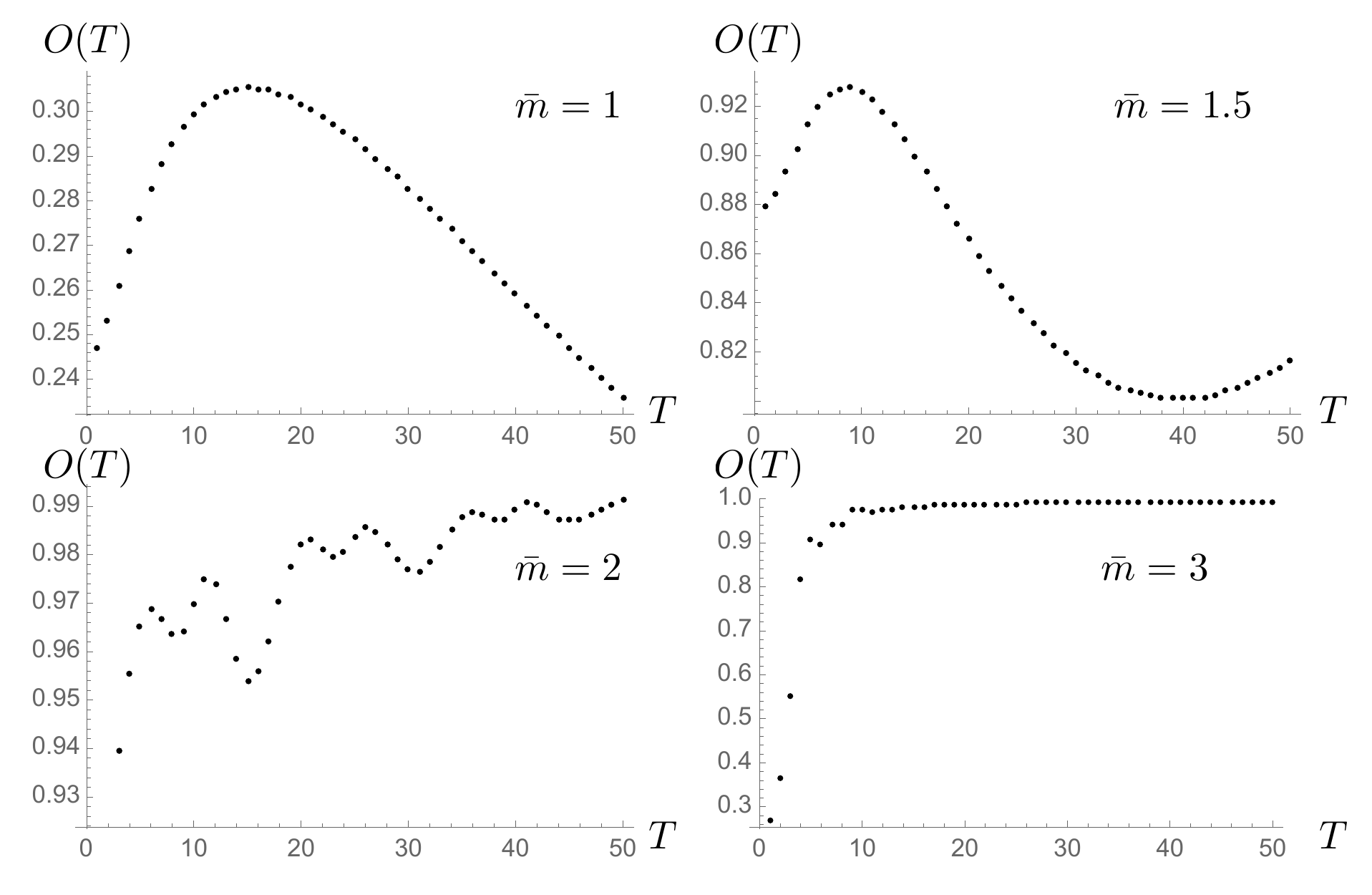} 
   \caption{The overlap of the wave function at time $T$ with the ground state (at appropriate mass) for SN.}
   \label{fig:olapplot}
\end{figure}

In~\cite{Meter}, the dynamics of SN is described as a ``partial collapse" to the ground state -- we can calculate the overlap of the wave function at time level $T$ with the ground state, $O(T) = |\langle \Psi(T)| \Psi_0 \rangle|$, and the plot of that overlap is shown in~\reffig{fig:olapplot}.  Notice that as the mass increases, the amount of overlap with the ground state increases.  For the lower masses, it is not clear what a longer temporal run would do (oscillate about some fixed, non-unity value, or increase towards full overlap), but for $\bar m = 2$ and $3$, a clear trend towards collapse to the ground state is shown.

\begin{figure}[htbp] 
   \centering
   \includegraphics[width=4in]{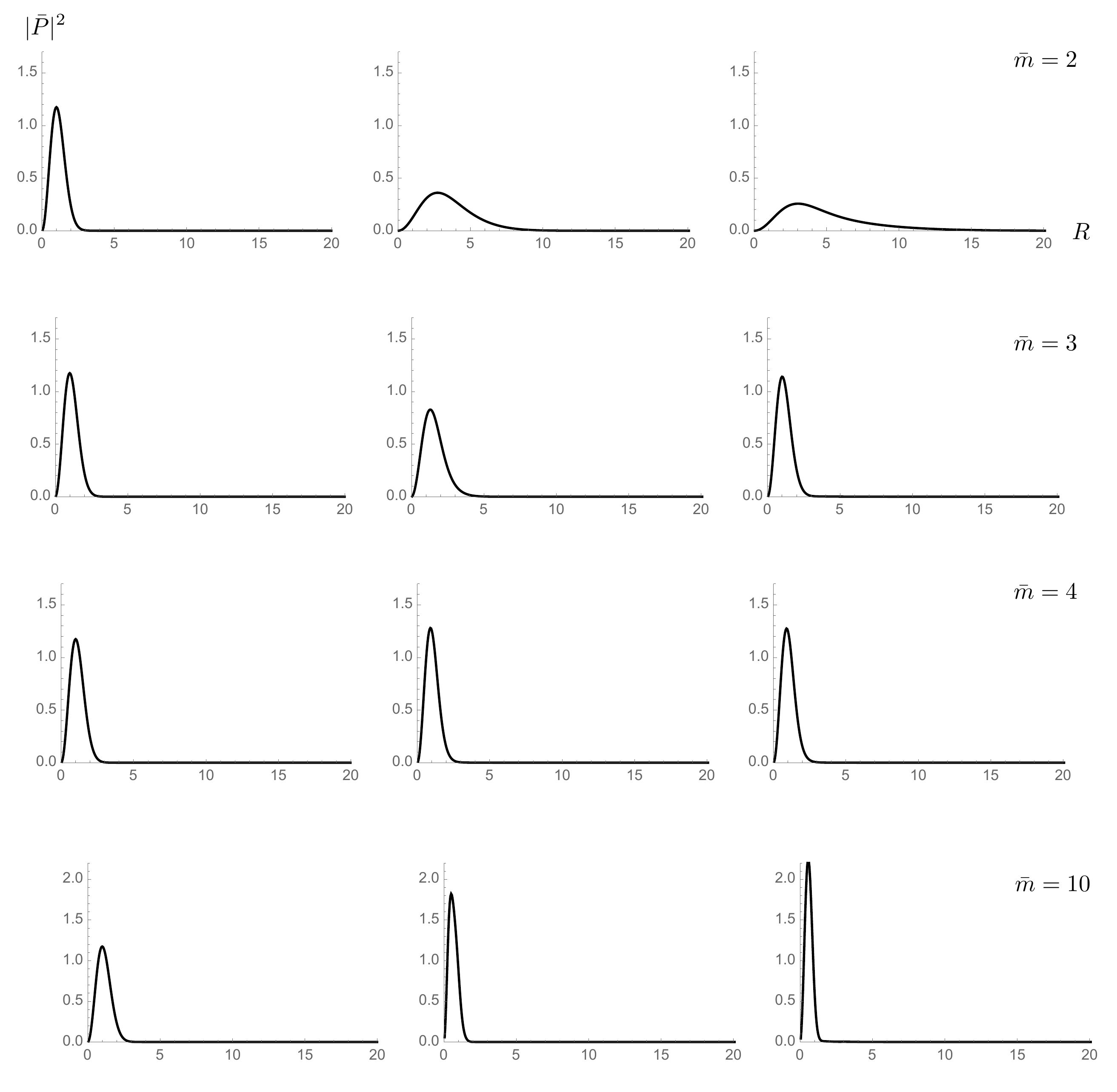} 
   \caption{Probability density as a function of position for self-coupled gravity masses $\bar m = 2$, $3$, $4$ and $10$.  Snapshots are shown at $T = 1 \,\Delta T$, $50 \, \Delta T$ and $100 \, \Delta T$  (left to right) in each case.  (Note the change in vertical scale).}
   \label{fig:SCmasses}
\end{figure}

\begin{figure}[htbp] 
   \centering
   \includegraphics[width=4in]{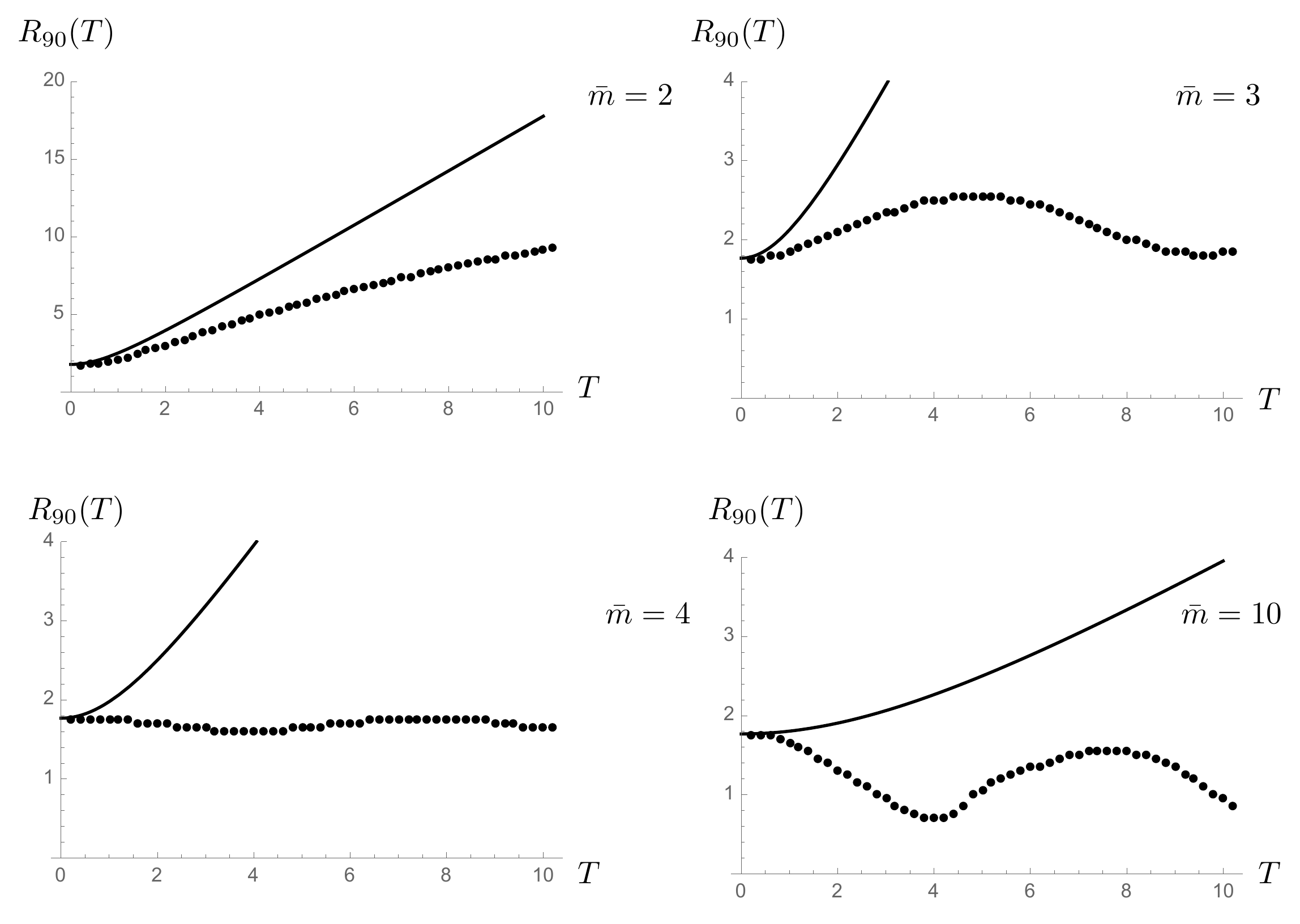} 
   \caption{The values of $R_{90}(T)$ for the self-coupled form of gravity at $\bar m =  2$, $3$, $4$ and $10$ are shown as points.  The line is the $R_{90}(T)$ for a free Gaussian.}
   \label{fig:SC90}
\end{figure}

Making the same plots for the self-coupled gravity case (with densities in~\reffig{fig:SCmasses} and $R_{90}(T)$ shown in~\reffig{fig:SC90}), at masses $\bar m = 2$, $\bar m = 3$, $\bar m = 4$ and $\bar m = 10$, we again see the spreading behavior at $\bar m =2$, and at $\bar m = 3$, oscillation has begun.  This oscillation does not represent collapse, though, as can be seen in~\reffig{fig:SC90}, the oscillation occurs at values {\it above} the initial $R_{90}(0)$ -- there is no contraction here.  It isn't until $\bar m =4$ that oscillation with values {\it below} the initial $R_{90}(0)$ occurs.  So we would put the critical mass somewhere between $\bar m = 3$ and $4$, again agreeing with our estimate $\sim 3.3$.  What is surprising in this case is the lack of decay we saw in, for example, $\bar m = 3$ of SN (both in the plot of $R_{90}(T)$ and in $O(T)$).   Instead, in the self-coupled case, all masses display oscillatory behavior without ``settling down" (we have run up to masses of $\bar m = 20$, but still see no sign of a collapse to the ground state).

\begin{figure}[htbp] 
   \centering
   \includegraphics[width=4in]{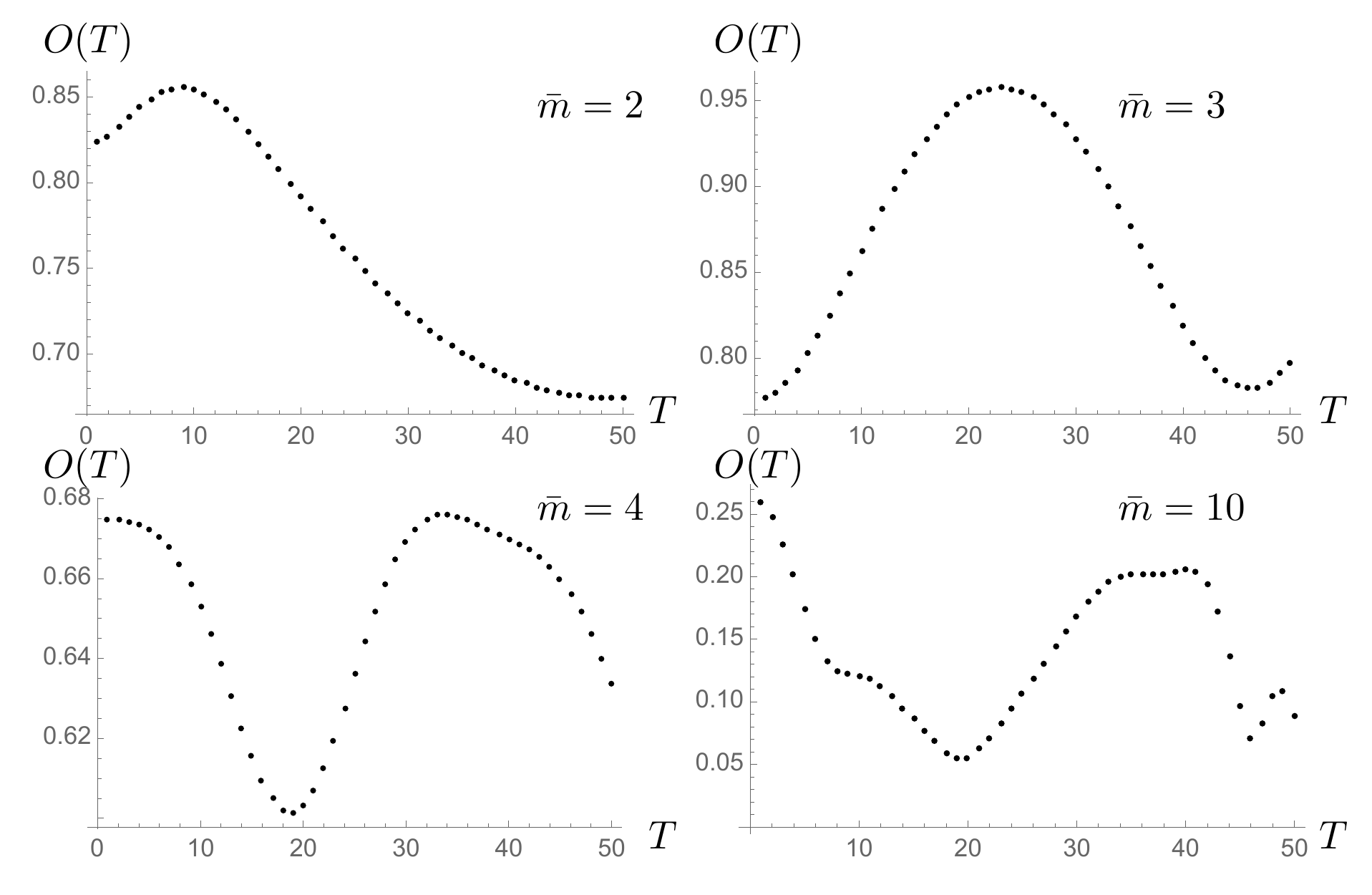} 
   \caption{The overlap of the wave function at time $T$ with the ground state (at appropriate mass) for the self-coupled case.}
   \label{fig:scolap}
\end{figure}

This lack of convergence can also be seen in the plots of the overlap with the ground state (calculated, appropriately, for the self-coupled case), shown in~\reffig{fig:scolap}.  Instead of oscillating towards an overlap of $1$ with the ground state, as in SN, 
  the overlap in the self coupled case does not increase (on average) over time (for the time scales considered here).  As another contrasting feature -- in~\reffig{fig:olapplot}, the amount of (time-averaged) overlap increases with mass, while in~\reffig{fig:scolap}, the magnitude of the overlap increases, but then decreases as mass gets larger.  

\section{Conclusion}

The inclusion of the self-coupling for gravity changes the spherical dynamics at large masses; while the expected qualitative behavior, free spreading and oscillation, occur in the expanded gravitational setting, the mass scales at which they occur are roughly twice those of Newtonian gravity.  We estimated the mass scales using a simple equivalence of quantum mechanical ``acceleration" and the gravitational field associated with our initial Gaussian wave function, and that estimate agreed fairly well with the numerical solutions.  The collapse to the ground state, apparent for SN at masses above $\bar m = 2$ here, is absent from the self-coupled case (at the time scales considered here -- time scales which are relevant for the SN case, at least).

Because we are using a form of gravity inspired by special relativistic mass-energy equivalence, we first calculated the energy spectrum of the quantum-mechanical/self-coupled gravitational system using the Dirac equation, to compare with the previously published Schr\"odinger spectrum, and found that, for the masses of interest to us at collapse, the error in the ground state energy is $\sim 10\%$, this suggests we can use the Schr\"odinger equation to evolve the initial Gaussian forward in time without incurring too much error.  For comparison, the difference between the ground state energy for SN and Dirac with self-coupled gravity is $\sim 600\%$.

Self-coupled gravity does not appear to collapse to its ground state (or any other); the wave function does not achieve a relatively static steady state, as it does in SN, nor does it ``converge" (in overlap) to its ground state.  It would be interesting to establish, analytically, that the ground state in the self-coupled form of gravity is dynamically unstable, leading to the observed oscillation without the decay to the ground state present in SN.  Another potential issue is our use of the Schr\"odinger equation -- perhaps at higher mass values, where the Dirac equation is relevant, we would find a damped-oscillatory collapse for the self-coupled gravity.


\begin{thebibliography}{18}
\bibitem{NOTE1} The Schwarzschild radius is, of course, external to either the Newtonian or self-coupled gravities considered here -- it represents an artificial length, from the point of view of the current work, and we understand that to truly describe the physics at this scale, additional gravitational elements (at the very least) are necessary.
\bibitem{NOTE2} While this simplified estimate works well for Newtonian gravity, using the point source solution for the modified, self-coupled gravity would yield no real $\bar m$ that satisfied~\refeq{cusp}, so we must use a more realistic distribution.
\end{thebibliography}
\end{document}